\documentclass[10pt,twocolumn]{paper}
\usepackage{epsfig, graphicx}
\usepackage[english]{babel}
\usepackage[margin=1.5cm]{geometry}

\title{\center \rm \bf Laboratory X-ray Study of the Phospholipid Monolayers at the Water Surface}

\author{\small \rm Viktor E. Asadchikov$^a$, Aleksey M. Tikhonov$^b$\/\thanks{tikhonov@kapitza.ras.ru}, 
Yurii O. Volkov$^a$, Boris S. Roshchin$^a$, Yurii A. Ermakov$^c$,\\
\small \rm Ekaterina B. Rudakova$^a$, Irina G. D'yachkova$^a$, and Alexander D. Nuzhdin$^a$\\
\small $^a$Shubnikov Institute of Crystallography, Federal Research Center Crystallography and Photonics,\\
\small Russian Academy of Sciences, Moscow, Russia\\
\small $^b$Kapitza Institute for Physical Problems, Russian Academy of Sciences, Moscow, Russia\\
\small $^c$Frumkin Institute of Physical Chemistry and Electrochemistry, Russian Academy of Sciences,
Moscow, Russia}

\begin{document}
\maketitle

\abstract{ \it The possibility of laboratory X-ray reflectometry study of the structure of dimyristoyl phosphatidylserine (DMPS) phospholipid monolayers on the water surface in various phase states has been demonstrated.}

\vspace{0.25in}

Studies of various phospholipid-based systems are of interest because of both fundamental aspects of
condensed matter physics and their basis role in biological membranes [1]. However, the preparation of
macroscopic samples of phospholipid bilayers or multilayers on solid substrates is limited because the radius of spontaneous curvature of a lipid bilayer in an aqueous medium is less than 50 $\mu$m [2]. For this reason, the X-ray and neutron small-angle scattering studies of the structure of the lipid bilayer in an aqueous medium were performed only for three-dimensional aggregates (vesicles) [3-6]. In view of this circumstance, X-ray studies of macroscopically planar monolayer and multilayer lipid structures on an extended horizontal surface of liquid substrates are certainly of interest.

X-ray reflectometry studies of the spatial structure of such samples are usually performed on specialized
synchrotron stations. The specificity of experiments for liquid samples (in particular, the necessity of the
horizontal arrangement of a sample) significantly complicates the design of an optical system and the
deflector system of a synchrotron beam. As a result, the number of stations equipped for the study of interfaces between liquids is comparatively small and their work load is high. Furthermore, the intensity of the synchrotron beam is high enough to induce the degradation of lipid films in a time comparable with the
duration of a single measurement of the angular dependence of reflected radiation [7].

We created an X-ray diffractometer with the horizontal arrangement of the sample and a mobile emitter–
detector system [8]. Such a design of the instrument allows X-ray reflectometry studies of liquid samples.
The possibility of studying the structure of phospholipid multilayers deposited on a liquid silica
sol substrate with this instrument was demonstrated in our works [9, 10]. However, it is noteworthy that the
formation of phospholipid monolayers and their structure on the water surface (which possibly better
simulates biological membranes) differ from those studied in our previous works. 

Moreover, the contrast in X-ray experiments is to significant extent determined by the ratio between the electron densities of the film and substrate. For example, according to [6], this ratio of the densities of the lipid mesophase and water lies in the range of $0.95 - 1.05$. For such low-contrast systems, a significant change in reflection and scattering curves can be comparable in order of magnitude to the experimental error of a detected signal. Thus, reflectometry study of lipid layers on water imposes additional requirements on the level of parasitic noise of an instrument.

\begin{figure}
\hspace{0.3in}
\epsfig{file=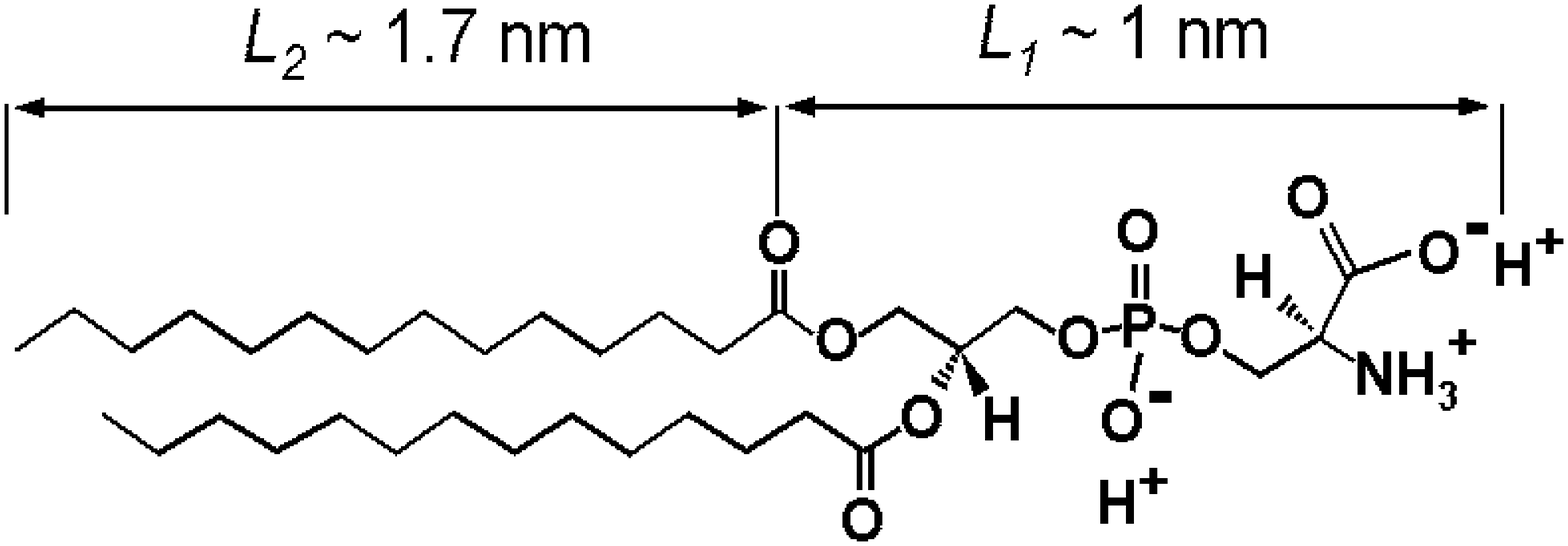, width=0.40\textwidth}

\small {\bf Figure 1}. Molecular structure of dimyristoyl phosphatidylserine
(DMPS).

\end{figure}

It is also worth noting that the solution of the inverse problem of X-ray reflectometry, i.e., the
reconstruction of the density of the structure of the sample in the direction perpendicular to its surface,
becomes better defined at a wider range of sliding angles in the experiment. In this case, the reflected
signal decreases naturally. In similar experiments performed on synchrotron stations, reflected radiation
can be detected at a decrease in its intensity by eight to ten orders of magnitude with respect to the primary beam. In this work, we show by example of the study of dimyristoyl phosphatidylserine (DMPS, see Fig. 1) monolayers on water substrates that results comparable in quality can also be obtained in laboratory experiments. Our setup and developed methods for data analysis allow detecting structural effects in thin planar layers on the surface of liquids, in particular, in the presence of surfactants,
e.g., phospholipids. The reported results make it possible to estimate geometrical factors
reflecting changes in DMPS in phospholipid molecules at a phase transition in the monolayer on the
water surface from the "expanded liquid" state to the gel, liquid-crystal state.

DMPS phospholipid monolayer samples were prepared and studied at room temperature $T\approx 298$\,K in an air-tight cell with X-ray transparent windows according to the method described in [10, 11]. A calibrated volume of the phospholipid solution in chloroform-methanol $5 : 1$ mixture was deposited by means of a microsyringe on the surface of the liquid substrate (KCl solution in deionized water at pH$\approx 7$) placed in a Teflon dish with the diameter $D=100$\,mm. The concentration of lipid in this solution was 0.5 mg/mL. In this work, we analyze data obtained for two 10 and 100\,mmol/L DMPS monolayers on the surface of the background KCl electrolyte. For the first and second samples (samples $a$ and $b$, respectively), the calculated specific area per molecule in the monolayer was $A = 100$ and $46$\,\AA$^2$, respectively. According to the previously studied dependence of the surface pressure $\Pi(A)$, monolayer $a$ is in the expanded liquid state, whereas monolayer $b$ is apparently a spatially inhomogeneous structure consisting of an equilibrium mixture of domains of the liquid and gel phases [12-15].

The angular dependence of the intensity of reflected radiation was measured in two stages. At the
first stage, the tube regime was chosen such that the intensity of reflected radiation was no more than
10$^4$\,pulse/s in order to avoid miscounts of the detector. As the angle increases and the intensity decreases below 10\,pulse/s, the tube regime was established at maximum values and the measurement was continued with the overlap of the angular range at the preceding stage by 0.1$^\circ$. The intensity of the incident beam at the maximum power on the tube is $\sim10^6$\,pulse/s and the characteristic background of the detector is 0.1\,pulse/s. Thus, the range of measurements reaches seven orders of magnitude in drop of the signal intensity.

At mirror reflection, the scattering vector {\bf q = k$_{\rm in}$ {\rm -} k$_{\rm sc}$}, where {\bf k}$_{\rm in}$ and {\bf k}$_{\rm sc}$ are the wave vectors of the incident and scattered beams in the direction to the observation point, respectively, has only one nonzero component $q_z=(4\pi/\lambda)\sin\alpha$, where $\alpha$ is the grazing angle in the plane normal to the surface (see the
inset of Fig. 2). The angle of total external reflection for the water surface $\alpha_c\approx\lambda\sqrt{r_e\rho_w/\pi}$$\approx0.15^\circ$ ($q_c=(4\pi/\lambda)\sin\alpha_c$ $\approx0.022$\,\AA$^{-1}$) is determined by the volume electron density in it $\rho_w\approx 0.333$  {\it e$^-$/}{\AA}$^3$, where $r_e = 2.814\cdot10^{-5}$ {\AA} is the classical radius of the electron.

An X-ray beam was prepared on the X-ray diffractometer used in this work by means of a three-slit collimation system and a Si(111) single reflection monochromator crystal. In order to increase the total intensity
of the X-ray beam, we used a broad-focus tube ($12\times 2$\,mm) with a copper anode. Owing to a larger diameter of the heating wire of the tube, its maximum allowed power is 20\% higher, but the intensity is distributed over a larger area. As a result, the width of the incident beam increases and a geometrical factor should be introduced. The monochromator crystal was tuned to the $K_{\alpha_1}$ line (the energy of photons $E \approx 8048$\,eV and the wavelength $\lambda = 1.5405 \pm 0.0001$\,\AA{}). The width of the beam is the width of the intensity distribution at the level an order of magnitude below the maximum. In our case, $d$ was about $0.55$\,mm. An estimate $s = d/\tan\alpha$ of the size of the illuminated region shows that the region illuminated by the X-ray beam at the first stage of measurements of the angular dependence of the intensity of reflected radiation at this $d$ value and angles $\alpha \sim \alpha_c$ is much larger than the
area of the surface of the sample.

For the exact correction of distortions of the measured dependence, before each experiment, we
recorded the profile of the direct beam $I_0(\beta)$, where $\beta$ is the angular position of the detector in the vertical plane in the absence of the sample at a fixed position of the source $\alpha = 0$.

The linear size of the beam in the vertical plane along the $Oz$ axis  is determined by $l_0$ the distance from the axis of rotation ($l_0\approx 570$\,mm). Let $z = l_0\tan\beta$. The length $l$ of the projection of the sample on the plane perpendicular to the direction of propagation of the beam is $l = L \sin\alpha$, where is $L \approx D$ the size of the sample along the beam.

The geometrical factor is introduced as the ratio of the total intensity of the entire incident beam to the
fraction of the intensity of its section appearing within the surface of the sample under the assumption that the maximum of the beam is at the center of the sample:
\begin{equation}
  \Pi(\alpha) = \frac{\int\limits_{-d}^{d}I_0(z)dz}{\int\limits_{-l/2}^{l/2}I_0(z)dz}.
\end{equation}

Finally, the specular reflection coefficient including the geometrical factor is $R(\alpha) = I(\alpha)/[I_0 \Pi(\alpha)]$, where $I(\alpha)$ and $I_0$ are the intensities of the reflected and incident beams, respectively.

The angular dependence of the reflection coefficient thus obtained can be considered in the approximation
of infinite length of the sample in the lateral direction. The inclusion of the dependence $\Pi(\alpha)$ gives a correction of  $\sim 8\%$ and less than 2\% to the parameter $R$ at $\alpha = 2\alpha_c$ and $\alpha = 3\alpha_c$, respectively.

At the second stage, beginning with angles $\alpha > 8 \alpha_c$, the intensity of reflection is measured not only at the grazing angle $\alpha$ but also at the angles $\alpha - \Delta\beta$ and $\alpha + \Delta\beta$, where the shift $\Delta\beta$ is the double angular width of the reflected beam and is $\approx 400^{\prime\prime}$. This is necessary for discriminations of the contribution of the parasitic scattering background in the bulk to the intensity measured by the detector. The resulting reflected intensity $I^\prime(\alpha)$ is calculated from three values $I(\alpha)$ by the formula $I^\prime(\alpha)=I(\alpha) - (I(\alpha- \Delta\beta) + I(\alpha+ \Delta\beta))/2$. Thus, $R(\alpha)=I^\prime(\alpha)/I_0$.

The software of the diffractometer makes it possible to specify a variable angular step, the width of the
slit of the detector, and the time of exposure, which allows optimizing the measurement of the reflection
coefficient $R$ decreasing rapidly with increasing $\alpha$. In the measurements, a step of the variation of $\alpha$ is determined by the character of the measured dependence and is usually varied within a range of $10^{\prime\prime}$ to $500^{\prime\prime}$.

The described approach to the inclusion of the background at large angles and the geometrical factor at small angles makes it possible to obtain the angular dependence of the reflection coefficient of X-rays in
the range of 1 to $10^{-8}$, which is comparable with results obtained on modern synchrotron stations [11, 16-21]. Such a result is certainly achieved not only by improving the method of experiment but also by increasing its time, which does not nevertheless exceed 10\,h.

The resulting experimental dependences of the reflection coefficient are shown in Fig. 2. The curve $R(q_z)$
for sample $a$ noticeably differs from the dependence for sample $b$ and both dependences include pronounced extrema. The latter property directly indicates the nonuniformity of the distribution of the reflecting density over the depth of the near-surface layer.

The determination of this distribution (i.e., the solution of the inverse problem) was performed by two
different methods. The first (model) method involves data on the structure of the molecule of the studied
lipid. It is known that the molecule of the studied lipid consists of a dense "head" (phosphatidylserine group) and less dense hydrocarbon tails. In the process of liquid - gel transitions, hydrocarbon tails are ordered and the head part is dehydrated, which also leads to a change in its electron density.

For this reason, the lipid monolayer is reasonably simulated in the form of a bilayer structure on the
water surface with smooth interfaces [22]:
\begin{equation}
\begin{array}{c}
\displaystyle
\rho=\frac{1}{2}\rho_{0}+\frac{1}{2}\sum_{j=0}^2(\rho_{j+1}-\rho_j)
{\rm erf}\left(\frac{l_j}{\sigma_0\sqrt{2}}\right),
\\ \\
\displaystyle
l_j=z+\sum_{n=0}^{j}L_n,
\\ \\
\displaystyle
{\rm erf}(x)=\frac{2}{\sqrt{\pi}}\int_0^x\exp(-y^2)dy,
\end{array}
\end{equation}
where $\rho_0 \equiv \rho_w$; $L_0 \equiv 0$ is the position of the (water-polar-group layer) interface $(z=0)$; $L_1$ ($\rho_1$) and $L_2$ ($\rho_2$) are the thicknesses (electron densities) of the polar groups of phosphatidylserine and hydrocarbon tails, respectively; and $\rho_3\approx 0$ is the bulk electron density in air.

\begin{figure}
\hspace{0.1in}
\epsfig{file=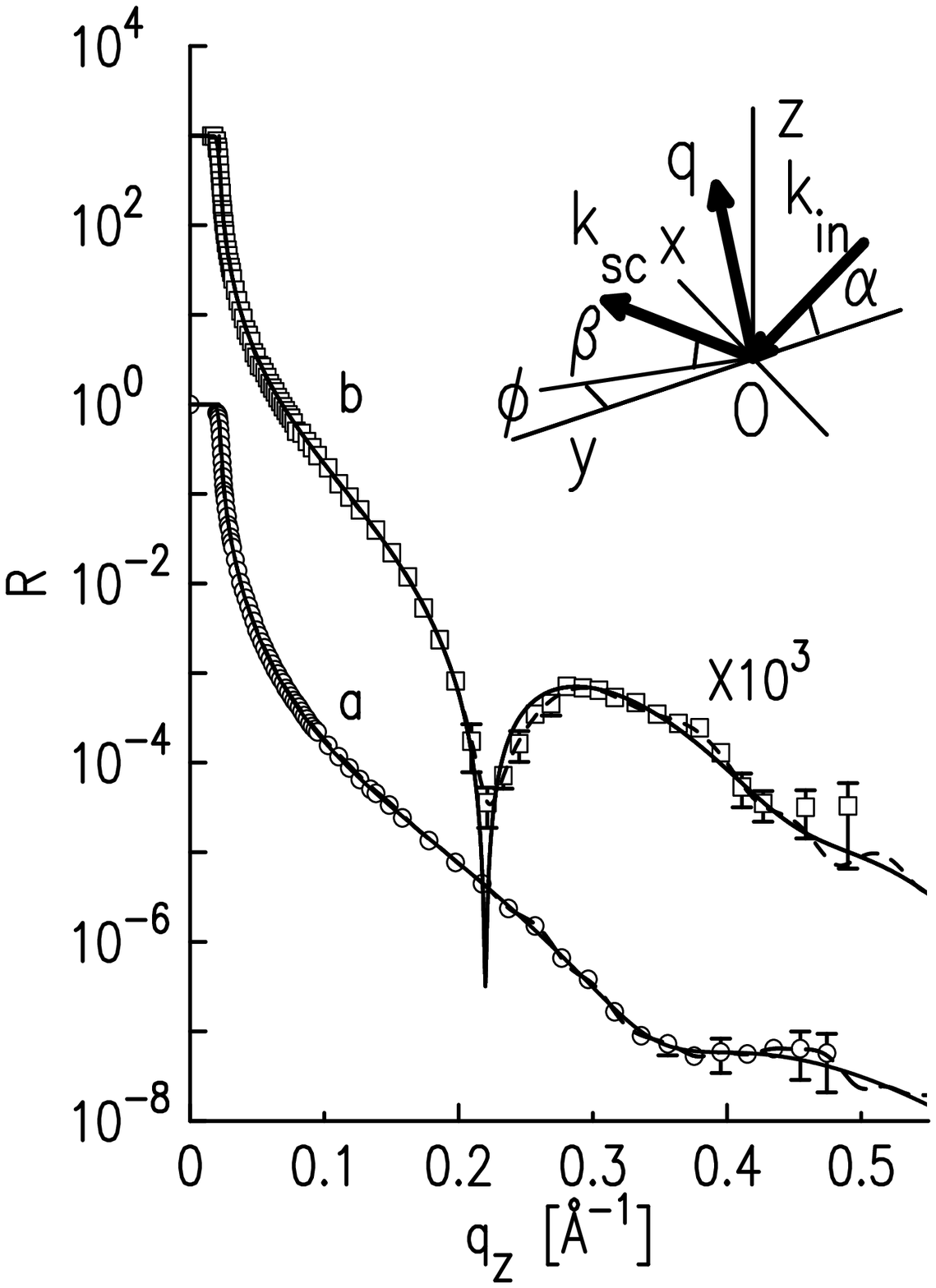, width=0.4\textwidth}

\small {\bf Figure 2}. Dependence $R(q_z)$ for the DMPS monolayer on the water surface for various areas per molecule: (circles) $A = 100$\,\AA$^2$ at the concentration of KCl in the substrate 10\,mmol/L and (squares) $A = 46$\,\AA$^2$ at the concentration of KCl in the substrate 100\,mmol/L. The solid lines correspond to the two-layer model of the monolayer given by Eq. (2), whereas the dashed lines are the results of the
free-form approach at the reconstruction of electron density profiles. The difference between these two approaches becomes noticeable at large sliding angles. The inset shows the kinematics of scattering in the coordinate system where $xy$ the plane coincides with the interface between the monolayer and water, $Ox$ the axis is perpendicular to the direction of the beam, and the $Oz$ axis is perpendicular to the surface and is opposite to the gravitational force. In the reflectometry experiment, $\alpha=\beta$ in the vertical $zy$ plane and $\phi=0$ in the horizontal plane.

\end{figure}

The parameter $\sigma_{0}$ determining the width of interfaces was fixed equal to the "capillary width"
$\sigma_{0}^2 = ( k_BT/2\pi\gamma ) \ln(Q_{max}/Q_{min})$ ( $k_B$ is the Boltzmann constant and $\gamma = 60 - 70$\,mN/m is the surface tension of the substrate), which is specified by the short-wavelength limit in the spectrum of capillary waves $Q_{max} = 2\pi/a$ ( $a\approx 10$\,{\AA} is the order of magnitude of the
intermolecular distance) and $Q_{min}=q_z^{max}\Delta\beta $( $2\Delta\beta$$\approx 1.7 \times 10^{-3}$\,rad is the angular resolution of the detector and $q_z^{max} \approx 0.5$ {\AA}$^{-1}$) [23-28]. In this representation, the theoretical value $\sigma_0$ for the chosen $A$ values is $2.7 - 3.2$\,\AA. The reflection coefficient $R(q_z)$ of X rays from the two-layer model thus specified can be easily calculated, e.g., in the distorted wave Born approximation [29]. Then, the desired structure was found by minimizing discrepancy between the calculated curve and experimental data with the thicknesses and electron densities of both parts of the lipid layer model as fitting parameters.

The calculation of the reflection coefficient $R(q_z)$ and the fitting of the parameters of the model profile were performed with the C-PLOT software (Certified Scientific Software) with one of the standard functions.
In this case, errors in the determination of the model parameters can be established with the use of
the standard criterion $\chi^2$.

The second approach is based on the extrapolation of the asymptotic behavior of the reflection curve $R(q_z)$ to the region of large$(q_z)$  values without any a priori assumptions on the transverse structure of the surface [30, 31]. This approach can be conventionally called "model-independent". It is assumed in this
approach that the depth distribution of the polarizability $\delta(z)$ contains singular points $z_j$ at which the polarizability (or its $n$th derivative) changes stepwise:
\begin{equation}
\Delta^{n}(z_j) \equiv  \frac{d^n\delta\left(z_j + 0\right)}{dz^n} - \frac{d^n\delta\left(z_j - 0\right)}{dz^n}.
\end{equation}
The set of such singular points unambiguously specifies the asymptotic behavior of the amplitude reflection
coefficient at $r(q_z)$ ïðè $q_z \to \infty$ ($R(q_z) \equiv \vert r(q_z) \vert ^2)$). The positions $z_j$ of the points can be determined from the experimental curve $R(q_z)$ measured in a limited range of $q_z$ values by means of the modified Fourier transform, which was described in detail in [30].

In the general case, there are only two physically reasonable distributions $\delta(z)$ that simultaneously satisfy the experimental values of the reflection coefficient $R(q_z)$ and a given set of singular points $\Delta^{n}(z_j)$ in the polarizability profile and differ only in the order of the positions of these points with respect to the substrate. 

The desired profile $\delta_z$ divided into $M \sim 100$ thin layers is described by a step function of the form $ \sum_{m=1}^{M}{\Delta^{n}(z_m)H(z-z_m)}$, where $H(z)$ is the Heaviside step function [32], with fixed positions of the singular points $\Delta^{1}(z_j)$. In turn, the reflection coefficient for such a structure $R(q_z,\delta(z_1)...\delta(z_M))$ can be calculated within the formalism of Parratt recurrence relations
[33]. The minimization of discrepancy between the calculated and experimental reflection curves, as
well as the fitting of the model profile $\delta(z_1...z_M)$, was performed in Python programming language environment with the use of the Scientific Python package, which implements the standard Levenberg–Marquardt
algorithm [34]. Finally, for weakly absorbed materials in a hard part of the X-ray spectrum, model-independent depth profiles of the electron density $\rho(z) \simeq \pi\delta(z)/(r_0\lambda^2)$ can be calculated from reconstructed distributions of the optical constant $\delta(z)$ [35].

Comparison indicates good agreement between the calculated and experimental reflection curves (see Fig. 2) for both samples. The electron density distribution in the lipid monolayer obtained with both reconstruction methods is shown in Fig. 3. It is seen that the chosen two-layer model of the structure is in good agreement with the electron density profile obtained independently within the model-independent (free-form) approach. This confirms the correctness of both the chosen monolayer model in general and the calculated parameters of its structural components.

For lipid film $a$, both the total thickness of the model structure $L_1+L_2 \approx 20$\,{\AA} and the distance $\approx 16${\AA} between singular points on the profile $\rho(z)$ in the model-independent approach are noticeably smaller the length of the lipid molecule $\approx 27$\,{\AA}. This fact indicates that hydrocarbon chains of molecular tails are disordered with respect to the normal to the surface.

According to the data for the second sample, the thickness of the second layer is $L_2 \approx 15$\,{\AA}, which approximately corresponds to the calculated length of 16.7\,{\AA} ($\approx  12\times 1.27$\,{\AA}(Ñ-Ñ) + 1.5\,{\AA}(-CH3)) of hydrocarbon tails -C$_{14}$H$_{27}$ in the DMPS molecule. The density $\rho_2$ and area per hydrocarbon chain $\approx 17$\,{\AA$^2$} in sample $b$ ($L_1+L_2 \approx 27$\,{\AA}) correspond to the crystal phase of a high-molecular weight saturated hydrocarbon [2]. The angle $\theta$ of deviation of the axis of molecular tails from the normal to the surface is $\theta = \arccos(15/16.7)\approx 30^\circ$. In turn, the thickness of the layer of polar heads $L_1$ is in the range of $10-12$\,{\AA} for both samples. The electron densities for samples $a$ and $b$ are $\sim 1.2\rho_w$ and $\sim 1.4\rho_w$, respectively. Such a difference is due to a change in the degree of hydration of polar groups of phospholipids at the compression of the monolayer [12].

\begin{figure}
\hspace{0.1in}
\epsfig{file=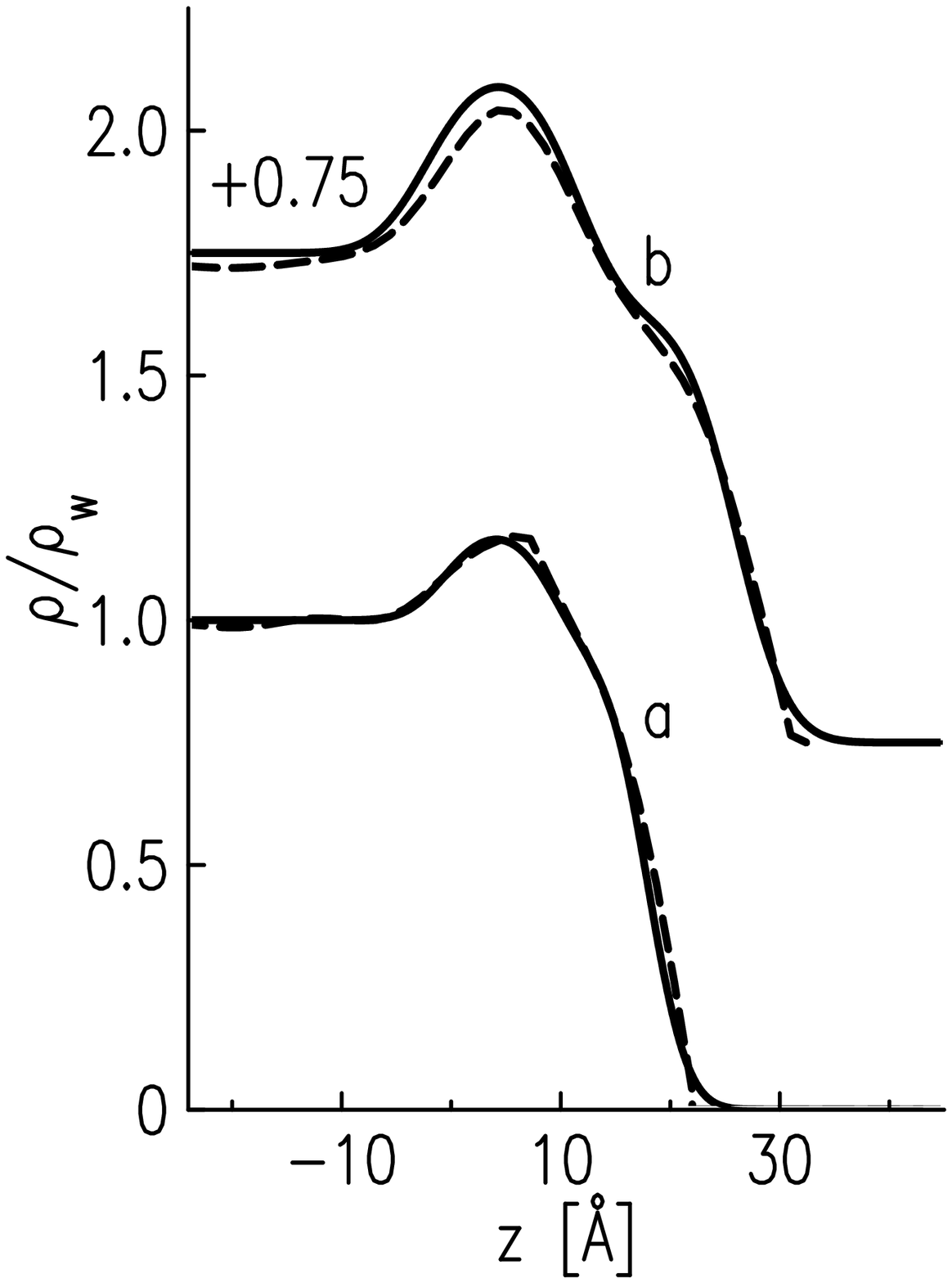, width=0.4\textwidth}

\small {\bf Figure 3}. Distribution profiles normalized to the electron density in water under normal conditions, $\rho_w\approx 0.333$ {\it e$^-$/}{\AA}$^3$: solid lines correspond to the "model" approach (see Eq. (2)), whereas the dashed lines correspond to the model-independent approach. For convenient comparison, lines for sample $b$ are shifted along the $y$ axis by 0.75 with respect to lines for sample $a$. The position of the interface between the polar region of lipid molecules and water is chosen at  $z = 0$.

\end{figure}

To summarize, we have experimentally justified the possibility of laboratory X-ray reflectometry study of
layers on water substrates. Data for the reflection coefficient $R(q_z)$ collected on our diffractometer are comparable in the spatial resolution $2\pi/q_z^{max} \approx 10$\,\AA{}  with the results previously obtained with synchrotron radiation. The methodical features of the experiment that made it possible to achieve such result have been described. Finally, data on the structure of DMPS lipid monolayers on a water substrate in various phase states have been obtained. 

We believe that the application of the described method of measurements to various monolayers, as well as our method of their analysis in combination with molecular dynamics calculations, makes it possible to reveal the features of the interaction of phospholipids with the water environment and the geometrical factors that significantly affect the distribution of electric fields in lipid membranes and near them [36].

\end{document}